\documentclass[aps,twocolumn,showpacs,eqsecnum,amsmath,amssymb]{revtex4}
\usepackage{graphicx}
\usepackage{dcolumn}
\usepackage{bm}
\begin{document}
\title{Observational Constraints on Two-field Warm Inflation}
\author{Yang-yang Wang}
\email{wangyy@mail.bnu.edu.cn}
\affiliation{Department of Physics, Beijing Normal University, Beijing 100875, China}
\author{Xiao-Min Zhang}
 \email{zhangxm@mail.bnu.edu.cn}
  \affiliation{School of Science, Qingdao University of Technology, Qingdao 266033, China}
\author{Jian-Yang Zhu}
\thanks{Corresponding author}
 \email{zhujy@bnu.edu.cn}
\affiliation{Department of Physics, Beijing Normal University, Beijing 100875, China}
\begin{abstract}
We study the two-field warm inflation models with a double quadratic potential and a linear temperature dependent dissipative coefficient. We derived the evolution equation of all kinds of perturbations without assuming slow-roll approximation, and obtained the curvature power spectrum at the end of inflation with a fully numerical method. Then we compute the scalar spectral index $n_s$, tensor-to-scalar ratio $r$ for several representative potentials, and compare our results with observational data. At last, we use Planck data to constrain the parameters in our models. This work is a natural extension of single-field warm inflation, and the aim of this work is to present some features of multi-field warm inflation using a simple two-field model.
\end{abstract}
\maketitle

\section{Introduction}

Inflation is widely accepted as the leading theory describing the early Universe \cite{inflationary universe a, a new inflationary, on the dynamics}, because it solves many long-standing puzzles of the hot big bang model, such as the horizon, flatness, and monopole problems. In addition, primordial fluctuations generated in inflation provide the seed for the large-scale structure of our Universe, and it can explain the temperature anisotropies on cosmic microwave background (CMB) naturally. In standard cold inflation, cosmological expansion and reheating are two separate period and we still know little about the details of reheating. The recent observational results have reached an impressive level of precision and improved the upper bound on tensor-to-scalar ratio $r$, so many representative models are ruled out in standard inflationary paradigm.

Warm inflation is an alternative to standard cold inflation \cite{warm inflation and}, where the interaction between inflaton and radiation can cause an extra dissipative term. The dissipative effects can lead to a sustainable radiation production, so the Universe can become radiation-dominated without reheating process. In warm inflation, density fluctuations come from thermal fluctuation, which is much larger than quantum fluctuations in cold inflation. Consequently, warm inflation can happen at a much smaller energy scale, and this leads to suppressed tensor-to-scalar ratio. Many inflationary potentials excluded in cold inflation become consistent with observations again in warm inflation. There have been various studies on warm inflation \cite{dynamic analysis of, adiabatic out-of-equilibrium solutions, consistency of the, identifying universality in}, and many interesting features are explored. Although warm inflation remains an appealing alternative to standard cold inflation, realizing warm inflation in concrete models is not easy. When coupling the inflaton directly with light fields, we have to make sure that the thermal corrections to the inflaton potential is not large, so that the inflaton potential remains flat and we can get sufficient e-foldings. At the same time, the dissipative effects should be strong enough to sustain a thermal bath at temperature $T>H$ during inflationary universe. In Ref.~\cite{is warm inflation}, the authors examined the feasibility of warm inflation from various viewpoint, and showed that it was extremely difficult or perhaps even impossible to realize the idea of warm inflation. Recently, a scenario fulfilling these conditions has been proposed in \cite{warm little inflation}, where the inflaton is a pseudo-Goldstone-boson coupled to a pair of fermionic fields through Yukawa interactions. In this case, the inflaton's mass gets protection from large thermal corrections due to the symmetries obeyed by the model, so the slow-roll of warm inflation will not be affected. This leads to enough dissipation with only a small number of fields, and a linear T dissipative coefficient.

The simplest inflationary models comprise only a single scalar degree of freedom, which are sufficient to obtain predictions consistent with observational constraints. Single-field inflation may seem natural from the perspective of simplicity and economy, but the status is not certain from perspective of microphysical origin of inflation \cite{curvature and isocurvature}. In fact, fundamental physics seems to predict the existence of a large number of scalar fields \cite{cosmology}. If the inflationary scale is not widely separated from the next relevant mass scale, it is natural to expect inflationary models with more than one active scalar fields. Inflation driven by multiple scalar fields has some specific features, such as large non-Gaussianity \cite{non-gaussianities in two-field, the next non-Gaussianity, primordial non-gaussianity in}, the presence of isocurvature perturbation \cite{the evolution of, on reaching the}, which can be narrowly constrained by the improved observational data in the future.

The content of the early universe is usually treated as a mixture of radiation fluid and scalar fields, in which interactions between different fields and dissipative effects play an important role. Therefore, to study the dynamical and perturbation features of multi-component cosmology and how it is constrained by observation is an important topic. One of the models including all these effects is the multi-field warm inflation paradigm, in which the radiation is regarded as a perfect fluid. In this paper we will focus on two-field warm inflation and try to reveal some features of multi-field inflation by constraining our models with observation data. In our previous work \cite{two-field warm inflation}, we have shown some simple conclusions in a temperature($T$) independent dissipative coefficient $\gamma$, which is not a realistic case. Now we extend our analysis as a next step to a linear $T$ dissipative coefficient. In Ref.~\cite{warm little inflation} it has been shown that a dissipative coefficient $\gamma =C_T T$ can be realized in a Little Higgs model, which put warm inflation on a solid footing in the aspect of model building.

This work is organized as follows. In Section II, we introduce some basics of two-field warm inflation, and derive the full set of equations describing the background dynamics and perturbations. In Section III, we give the main features of two-field warm inflation by numerically solving the equations obtained in Section II. In Section IV, we use the Planck data to constrain our models, and give the observational constraints on model parameters. In Section IV, we present our conclusions.

\section{Basics of Two-field Warm Inflation}
The two-field warm inflation dynamics is characterized by the coupled background equations of inflaton field $\phi(t)$, $\chi(t)$ and radiation density $\rho_r(t)$,
\begin{gather}
\ddot{\phi}+\left( 3H+\gamma \right) \dot{\phi}+V_{\phi}=0, \nonumber \\
\ddot{\chi}+\left( 3H+\gamma \right) \dot{\chi}+V_{\chi}=0, \nonumber \\
\dot{\rho_r}+4H \rho_r = \gamma \left( {\dot{\phi}}^2 + {\dot{\chi}}^2 \right), \label{background equations}
\end{gather}
where $V$ is the inflaton potential, $V_{\phi} = \partial V(\phi, \chi)/\partial \phi$, $V_{\chi} = \partial V(\phi, \chi)/\partial \chi$, overdots represent derivatives with respect to cosmic time $t$, and $\gamma$ is dissipative coefficient. In general case, $\gamma$ is a function of background inflaton fields and the temperature $T$.
For simplicity, we will use the Planck Unit in the following context,
$$8\pi G=k_B=\hbar=c=1,$$
where $G$ is Newton's gravitational constant, $k_B$ is Boltzmann's constant,
$\hbar$ is the reduced Planck's constant, and $c$ is the speed of light. In a spatially flat universe, the Friedmann equations read:
\begin{equation}
3H^2=\frac{1}{2} {\dot{\phi}}^2+\frac{1}{2} {\dot{\chi}}^2+V(\phi, \chi) + \rho_r, \label{FRW}
\end{equation}
and the slow-roll parameters are defined as,
\begin{equation}
\epsilon=-\frac{\dot{H}}{H^2}, \quad \eta=-\frac{\ddot{H}}{2H \dot{H}}. \label{slow-roll parameters}
\end{equation}

Inflation takes place when the slow-roll condition $\epsilon<1$, $| \eta |<1$ are satisfied. In slow-roll approximation we have,
\begin{gather}
\left(3H+\gamma \right) \dot{\phi}+V_\phi=0, \\
\left(3H+\gamma \right) \dot{\chi}+V_\chi=0, \\
4H \rho_r= \gamma \left( \dot{\phi}^2+\dot{\chi}^2 \right).
\end{gather}

For convenience, we define adiabatic field $\sigma$ and entropy field $s$ by making a rotation in field space, where $d \sigma$ is tangent to the background trajectory and $d s$ is normal to it \cite{adiabatic and entropy, spectral running and},
\begin{gather}
\left( \begin{array}{c} d \sigma \\ d s \end{array} \right)=\left( \begin{array}{cc} \cos \theta \quad \sin \theta \\ -\sin \theta \quad \cos \theta \end{array} \right) \left( \begin{array}{c} d \phi \\ d \chi \end{array} \right), \label{background transform}
\end{gather}
where $\cos \theta=\frac{\dot{\phi}}{\sqrt{\dot{\phi}^2+\dot{\chi}^2}}$, $\sin \theta=\frac{\dot{\chi}}{\sqrt{\dot{\phi}^2+\dot{\chi}^2}}$. With this definition, the background equations Eqs.~\eqref{background equations} and \eqref{FRW} become,
\begin{gather}
\ddot{\sigma}+\left(3H+\gamma \right) \dot{\sigma}+V_\sigma=0, \nonumber \\
\dot{\theta} \dot{\sigma}+V_s=0, \nonumber \\
\dot{\rho_r}+4H \rho_r = \gamma \dot{\sigma}^2, \nonumber \\
3H^2=\frac{1}{2} \dot{\sigma}^2+V+\rho_r
\end{gather}
where $V_\sigma = \cos \theta V_\phi + \sin \theta V_\chi$, $V_s = -\sin \theta V_\phi+ \cos \theta V_\chi$. In this case, slow-roll parameters can be expressed by,
\begin{gather}
\epsilon=\epsilon_\sigma+\epsilon_r, \ \epsilon_\sigma=\frac{1}{2} \frac{\dot{\sigma}^2}{H^2}, \ \epsilon_r=\frac{2}{3} \frac{\rho_r}{H^2}, \nonumber \qquad \quad \\
\eta=\frac{\epsilon_\sigma}{\epsilon} \eta_\sigma + \frac{\epsilon_r}{\epsilon} \eta_r, \ \eta_\sigma=- \frac{\ddot{\sigma}}{H \dot{\sigma}},\ \eta_r=-\frac{1}{2} \frac{\dot{\rho_r}}{H \rho_r}. \label{slow-roll parameters2}
\end{gather}

In order to study the evolution of perturbations, we decompose each of the fields into a spatially homogenous background field and its perturbations, $\phi (x,t) \rightarrow \phi(t)+\delta \phi (x,t)$, $\chi(x,t) \rightarrow \chi(t)+\delta \chi(x,t)$. Similarly, it is convenient to decompose the field perturbation into an adiabatic component $\delta \sigma$ and entropy component $\delta s$ \cite{correlation-consistency cartography of},
\begin{gather}
\left( \begin{array}{c} \delta \sigma \\ \delta s \end{array} \right)=\left( \begin{array}{cc} \cos \theta \quad \sin \theta \\ -\sin \theta \quad \cos \theta \end{array} \right) \left( \begin{array}{c} \delta \phi \\ \delta \chi \end{array} \right). \label{perturbation transform}
\end{gather}

The line element of the Friedmann-Robertson-Walker (FRW) metric is given by,
\begin{multline}
d s^2 = - (1 + 2 A) d t^2 + 2 a \partial_i B d x^i d t \\
+ a^2 ((1 - 2 \psi) \delta_{ij} + 2 \partial_i \partial_j E) d x^i d x^j.
\end{multline}

In warm inflation, we have to take into account the perturbations result from both the inflation field and radiation. Therefore, the total gauge-invariant comoving curvature perturbation can be split into two parts \cite{oscillatory power spectrum, adiabatic and entropy2, cosmological perturbations in},
\begin{equation}
\mathcal{R}=\frac{\epsilon_\sigma}{\epsilon} \mathcal{R}_\sigma + \frac{\epsilon_r}{\epsilon} \mathcal{R}_r, \label{perturbation split}
\end{equation}
where $\mathcal{R}_\sigma = \psi + H \frac{\delta \sigma}{\dot{\sigma}}$, $\mathcal{R}_r = \psi - a H \left(B+\delta u \right)$, and $\delta u$ is the scalar velocity potential of the radiation fluid. For convenience, we also define the isocurvature perturbation $\mathcal{S}=H \frac{\delta s}{\dot{\sigma}}$. The power spectrum of comoving curvature $\mathcal{R}$ perturbation is given by,
\begin{equation}
\mathcal{P}_{\mathcal{R}}=\frac{k^3}{2\pi^2} \langle \mathcal{R}^2 \rangle, \label{total spectrum}
\end{equation}
and the power spectrum of inflaton perturbation $\mathcal{R}_\sigma$, radiation perturbation $\mathcal{R}_r$ and isocurvature perturbation $\mathcal{S}$ are \cite{on the importance},
\begin{equation}
\mathcal{P}_\sigma = \frac{k^3}{2 \pi^2} \langle \mathcal{R}_\sigma^2 \rangle, \ \mathcal{P}_r = \frac{k^3}{2 \pi^2} \langle \mathcal{R}_r^2 \rangle, \ \mathcal{P}_\mathcal{S} = \frac{k^3}{2 \pi^2} \langle \mathcal{S}^2 \rangle, \label{individual spectrum}
\end{equation}
where $k$ denotes comoving wave number. In warm inflation, density perturbations mainly arise from thermal noise \cite{non-gaussianity in fluctuations}. On small scales ($k \gg a H$) the metric fluctuations have little effects \cite{scalar perturbation spectra, a relativistic calculation}, so inflaton fluctuations $\delta \varphi_I$($\delta \varphi_I = \delta \phi, \delta \chi$) are described by a Langevin equation \cite{density fluctuations from},
\begin{equation}
\ddot{\delta \varphi_I}(k,t)+(3H+\gamma) \dot{\delta \varphi_I}(k,t)+\frac{k^2}{a^2} \delta \varphi_I = \xi_I(k,t),
\end{equation}
where $\xi_I(k,t)$ is a white-noise term and different components of $\xi_I(k,t)$ is independent of each other. In this case, there is no coupling between different components of inflaton perturbations before horizon-crossing, which is a common assumption in standard multi-field inflation. In the high temperature limit, the noise source is Markovian,
\begin{equation}
\langle \xi_I (k, t) \xi_J (- k', t') \rangle = 2 \gamma T a^{- 3} (2 \pi)^3 \delta_{I\!J} \delta^3 (k -
k') \delta (t - t'),
\end{equation}
where $T$ denotes the temperature and $a$ is the scales factor. The relationship between radiation energy density $\rho_r$ and $T$ is $\rho_r=\frac{\pi^2}{30} g_* T^4$, where $g_*$ is the effective particle number of radiation fluid \cite{observational constraints on warm}. We will take $g_*=228.75$ in the following numerical calculation, which is the number of degrees of freedom for the Minimal Supersymmetric Standard Model \cite{shear viscous effects, observational constraints on}. After horizon-crossing, the effects of thermal noise are suppressed while the metric perturbation come to play an important role \cite{cosmological inflation and}. In previous studies, the perturbations in sub-horizon and super-horizon scales are often treated separately for simplicity. The power spectrum in warm inflation has been studied in many previous works \cite{power spectrum for, exploring the parameter, evolution of the}, and a general expression for amplitude of inflaton power spectrum is given by \cite{constraining warm inflation},
\begin{equation}
\mathcal{P}_{\delta \sigma}^*=\left( \frac{H_*}{2 \pi} \right)^2 \left( \frac{T_*}{H_*} \frac{2 \pi Q_*}{ \sqrt{1+4 \pi Q_*/3}} +1+2n_*\right) G(Q_*), \label{analytic expression}
\end{equation}
where a subscript ``*'' denotes variables evaluated at horizon crossing, $Q=\gamma /(3H)$ is the dissipative ratio, and $n_*=1/\left(e^{{H_*}/{T_*}}-1\right)$ is the statistical distribution of inflaton fluctuations at horizon crossing \cite{dynamical and observational}. The function $G(Q_*)$ represents the growth of $\mathcal{P}_{\mathcal{R}}$ due to the coupling between inflaton fluctuations and radiation fluctuations, and this growing function can only be determined by solving the perturbation equations numerically.

When performing numerical calculations, it is more convenient to take e-foldings $N$ ($d N= H dt$) as the time variable. In order to get a full picture of the evolution of perturbations, we take into account all kinds of perturbations in two-field warm inflation, and give the evolution equation of $\mathcal{R}_\sigma$, $\mathcal{R}_r$ and $\delta s$ beyond slow-roll approximation (see Appendix A for more details),

\begin{widetext}
\begin{multline}
{\mathcal{R}_\sigma} ''+ \left(3+3Q+\epsilon_r+\epsilon_\sigma - 2 \eta_\sigma \right) {\mathcal{R}_\sigma}' + \left(z^2 -2{\lambda_\theta}^2 \epsilon_\sigma -9Q + 2 (1+\epsilon_r- \eta_\sigma) \epsilon_r +Q (3\epsilon_r-11 \epsilon_\sigma+6 \eta_r) \right) \mathcal{R}_\sigma \\
=- (3Q+2 \epsilon_r) {\mathcal{R}_r}' +\left(-9Q+2(1+\epsilon_r - \eta_\sigma) \epsilon_r + Q(3\epsilon_r - 11 \epsilon_\sigma +6 \eta_r) \right) \mathcal{R}_r \\
+2 \lambda_\theta {\delta s}' +2 \left(\lambda_\theta(3+3Q-\epsilon_r-\eta_\sigma) + \lambda_{\theta \theta} \right) \delta s + \frac{z^{3/2} k^{-3/2}}{\sqrt{2 \epsilon_\sigma}} \xi_\sigma, \label{inflaton perturbation equation}
\end{multline}
\begin{multline}
{\mathcal{R}_r}''+ \left(6+\epsilon_r+\epsilon_\sigma - \frac{7}{2} \eta_r \right) {\mathcal{R}_r}' \\
+ \left(\frac{1}{3} z^2+9+\frac{1}{2} (-9+\epsilon_\sigma) \eta_r +2 (4+\epsilon_\sigma -2 \eta_r) (\epsilon_r+\epsilon_\sigma -\eta_\sigma) + \left(-1+Q+ \frac{2}{3} \epsilon_r \right) \epsilon_\sigma \right) \mathcal{R}_r \\
= \left(\frac{2}{3} \epsilon_\sigma +\frac{5}{4} (4-2\eta_r) \right) {\mathcal{R}_\sigma}'
+ \left(9+\frac{1}{2} (-9+\epsilon_\sigma) \eta_r +2 (4+\epsilon_\sigma -2 \eta_r) (\epsilon_r+\epsilon_\sigma -\eta_\sigma) + \left(-1+Q+ \frac{2}{3} \epsilon_r \right) \epsilon_\sigma\right) \mathcal{R}_\sigma \\
+ \frac{1}{3} (-6+2 \epsilon_\sigma +3 \eta_r) \lambda_\theta \delta s, \label{radiation perturbation equation}
\end{multline}
\begin{multline}
{\delta s}''+ (3+3Q-\epsilon_r - \epsilon_\sigma) {\delta s}' + \left( z^2+\frac{V_{ss}}{H^2} - 2 {\lambda_\theta}^2 \epsilon_\sigma \right) \delta s \\
= -4 \lambda_\theta \epsilon_\sigma {\mathcal{R}_\sigma}' - 4 \lambda_\theta \epsilon_\sigma \epsilon_r \mathcal{R}_\sigma + 4 \lambda_\theta \epsilon_\sigma \epsilon_r \mathcal{R}_r + z^{3/2} k^{-3/2} \xi_s, \label{isocurvature perturbation equation}
\end{multline}
\end{widetext}
where a prime denotes a derivative with respect to e-folding $N$, and $\lambda_\theta=\theta '/\sigma '$, $\lambda_{\theta \theta}=\theta ''/\sigma '$. $V_{ss}=\sin^2 \theta V_{\phi \phi}+ \cos^2 \theta V_{\chi \chi}$ is the effective mass of entropy field $s$, and $z=\frac{k}{aH}$. $\xi_\sigma$, $\xi_s$ are two gaussian white noise and their correlation function are,
\begin{equation}
\langle \xi_I (k, N) \xi_J (-k', N') \rangle=2 \gamma T (2 \pi)^3 \delta_{IJ} \delta^{(3)}(k-k') \delta(N-N')
\end{equation}
where $I, J=\sigma, s$. Similarly, using $N$ as the time variable, we can put Eqs.~\eqref{background equations} and \eqref{FRW} in the form,
\begin{gather}
\phi ''+(3+3Q-\epsilon) \phi ' +\frac{V_\phi}{H^2} = 0, \label{be1} \\
\chi '' + (3+3Q - \epsilon) \chi ' + \frac{V_\chi}{H^2}=0, \label{be2} \\
\rho_r ' + 4 \rho_r = 3H^2 Q \left( {\phi '} ^2 +{\chi '} ^2 \right), \label{be3} \\
\qquad \left(3- \frac{1}{2} {\phi '}^2- \frac{1}{2} {\chi'}^2 \right) H^2=V+\rho_r. \label{background equations2}
\end{gather}

Note that the slow-roll parameters in above equations are treated as a function of time variable $N$, and can be determined by solving background equations Eqs.~\eqref{be1}, \eqref{be2}, \eqref{be3} and \eqref{background equations2}. When dealing with multi-field inflation, it is necessary to go beyond slow-roll, or we will miss some important features. In this case, the numerical method is almost essential, because we can hardly find any analytic results.

\section{Numerical Examples}
In order to get a picture of the dynamics and perturbations of two-field warm inflation, we apply the formalism to a simple example in this section. We use the two-field quadratic inflation as an example, in which the potential is given by \cite{double inflation and},
\begin{equation}
V(\phi, \chi) = \frac{1}{2} m_\phi^2 \phi^2 + \frac{1}{2} m_\chi^2 \chi^2. \label{inflaton potential}
\end{equation}

Although these two scalar fields have no direct coupling, they can interact gravitationally during inflation. We will show that even the simplest two-field warm inflation models can display interesting features, which is very different from the single-field cases. In the following calculations, we set $N=0$ when the relevant scales cross the Hubble horizon, and we fix $N_e=60$ when inflation ends for definite calculation.

\begin{center}
\begin{figure}
\includegraphics[width=0.37\textwidth]{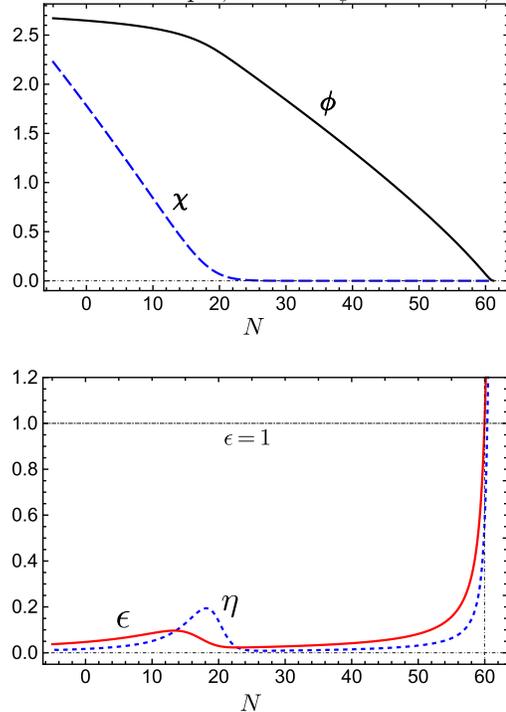}
\caption{The evolution of inflaton $(\phi, \chi)$ and slow-roll parameters $\epsilon$, $\eta$ are shown against e-foldings $N$. In this example, we choose the initial condition $(\phi_0, \chi_0)=(2.67, 2.23)$, $C_T=0.048$, and inflation ends at about $N=60$ in this case. From the lower panel, we can know that when the field $\chi$ decay to zero, a local extreme occurs for slow-roll parameters $\epsilon$ and $\eta$.}
\label{Fig-1}
\end{figure}
\end{center}

At the beginning of our analysis, we use a representative example to demonstrate the main features of background dynamics and perturbations of two-field warm inflation. In the example, we set $m_\phi=2 \times 10^{-8}$, and the mass ratio $R_{\rm m}=m_\chi/m_\phi=5$. After choosing the initial condition $(\phi_0, \chi_0)=(2.67, 2.23)$ and $C_T=0.0483$, we can perform our numerical computation by solving the background equations Eqs.~\eqref{be1}, \eqref{be2}, \eqref{be3}, \eqref{background equations2} and stochastic perturbation equations Eqs.~\eqref{inflaton perturbation equation}, \eqref{radiation perturbation equation} and \eqref{isocurvature perturbation equation}. Note that since the background dynamics will tend to the slow-roll trajectory soon, the initial value of $\dot{\phi}_*$, $\dot{\chi}_*$ and ${\rho_r}_*$ have little impact on the final results. In order to eliminate the influence of initial conditions of perturbations equations, we begin our numerical integration about 5 e-foldings before Hubble exit. The results are shown in Fig.~\ref{Fig-1} and Fig.~\ref{Fig-2}.

\begin{widetext}
\begin{center}
\begin{figure}
\includegraphics[width=0.85 \textwidth]{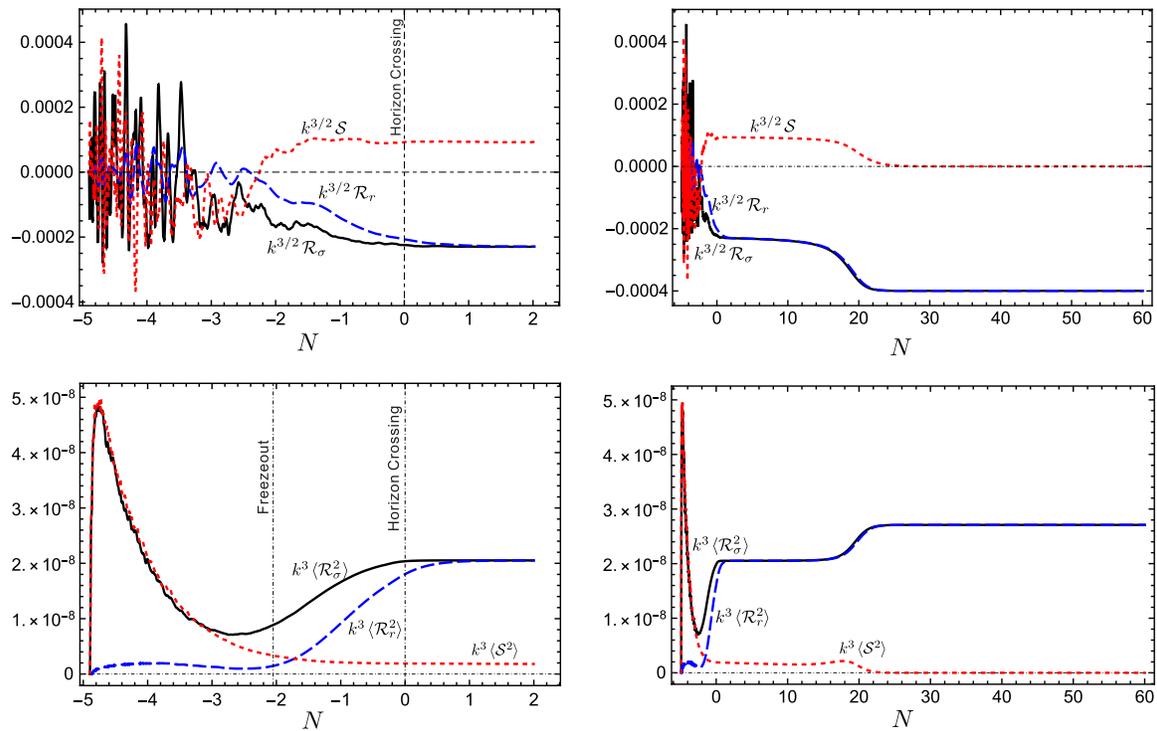}
\caption{The evolution of perturbations $\mathcal{R}_\sigma$, $\mathcal{R}_r$, $\mathcal{S}$, and the power spectrum of each kind of perturbations are shown in this figure. In the top two panels, we show a single realization of stochastic perturbation equation, and the power spectrum of each perturbation are illustrated in the lower two panels, which represent an average over 5000 runs. We zoom in to the few e-foldings around horizon crossing to given more details in the left two panels. }
\label{Fig-2}
\end{figure}
\end{center}
\end{widetext}

In warm inflation, thermal effects decrease until the fluctuations freezes out which is determined by $k^2/a^2 \approx (3H+\gamma)H$, and the freezeout time $N_F$ always precedes the horizon crossing time \cite{noncanonical warm inflation}. According to Fig.~\ref{Fig-2} we know before freezeout time the system is dominated by stochastic noise, and the initial condition of perturbations have little impact on the final results. In the period between freeze-out and horizon crossing, the power spectrum $\mathcal{P}_\sigma$ and $\mathcal{P}_r$ may get enhanced due to the coupling between perturbations of radiation and inflaton, as illustrated in the lower left panels of Fig.~\ref{Fig-2}. This growing mode occurs only when $Q_*>1$ (in the example illustrated in Fig.~\ref{Fig-2}, $Q_* \approx 10$), and in weak regime of warm inflation this is replaced by a constant mode \cite{cosmological fluctuations of}. After horizon-crossing, the power spectrum $\mathcal{P}_{\mathcal{R}}$ do not change for a while until the turning occurs at background trajectory. When background evolution trajectory change direction, the parameter $\dot{\theta}^2$ can become large, which will make the the perturbation $\mathcal{R}_{\sigma}$, $\mathcal{R}_r$ coupled strongly with isocurvature perturbation $\mathcal{S}$ as shown in Eqs.~\eqref{inflaton perturbation equation}, \eqref{radiation perturbation equation} and \eqref{isocurvature perturbation equation}. At this moment, $\mathcal{R}_{\sigma}$, $\mathcal{R}_r$ increases obviously while $\mathcal{S}$ decays to zero. In the following period there is only one active scalar field and the curvature perturbation tend to a constant value. We also know from the top two panels of Fig.~\ref{Fig-2} that $\mathcal{R}_{\sigma} \approx \mathcal{R}_r$ on super-horizon scales, and they evolve together until the end of inflation. Therefore, we can use $\mathcal{R}_{\sigma}$ to represent curvature perturbation $\mathcal{R}$ according to Eq.~\eqref{perturbation split}.

\section{Constraints from Observational Data}

In previous subsection we have known the main features of background dynamics and perturbation power spectrum in two-field warm inflation, and now we turn to the consistency with observational data. In order to compare with the observational data, we should get the power spectrum $\mathcal{P}_{\mathcal{R}}$ at the end of inflation first, and then we can obtain the spectral index $n_s$ using finite difference method \cite{computing observables in},
\begin{equation}
n_s-1=\frac{d \ln \mathcal{P}_{\mathcal{R}}}{d \ln k}. \label{spectral index}
\end{equation}

The tensor mode of perturbations is not affected by the thermal noise, so the tensor power spectrum and tensor-to-scalar ratio are given by \cite{inflation and the},
\begin{equation}
\mathcal{P}_T=8\left(\frac{H_*}{2 \pi} \right)^2, \quad r=\frac{\mathcal{P}_T}{\mathcal{P}_{\mathcal{R}}}. \label{tensor-to-scalar ratio}
\end{equation}

 There exist two methods available to compute the final $\mathcal{P}_{\mathcal{R}}$. The most straightforward way is the method we use in the previous section, where we perform our analysis by solving the coupled stochastic system numerically until the end of inflation. However, this is a computationally intensive way and the calculations consumes a long CPU time, because we have to perform tens of thousands of runs to get a relative accurate result. There exists another approach to achieve our purpose which is used more widely. We can use the analytic expression Eq.~\eqref{analytic expression} to express $\mathcal{P}_{\delta \sigma}$ at horizon crossing, and the growing function $G(Q_*)$ can be determined by integrating the stochastic equations a few e-foldings before horizon crossing. After horizon crossing, we can use $\delta N$-formalism to get the time evolution of $\mathcal{P}_{\mathcal{R}}$ until the end of inflation,
\begin{equation}
\mathcal{P}_{\mathcal{R}}={\mathcal{P}}_{\delta \sigma}^* \left(N_{\phi_*}^2+N_{\chi_*}^2\right), \label{delta-N formulism}
\end{equation}

where ${\mathcal{P}}_{\delta \sigma}^*$ is the power spectrum of field perturbation $\delta \sigma$ at horizon crossing.

In the following investigation, we take four representative values of mass ratio $R_{\rm m}$, $R_{\rm m}=1$, $R_{\rm m}=1.5$, $R_{\rm m}=2$ and $R_{\rm m}=3$ as examples. To obtain the growing function $G(Q_*)$ in Eq.~\eqref{analytic expression} in two-field warm inflation, we carry out a numerical simulation for each value of $R_{\rm m}$. $G(Q_*)$ and numerical results of simulations are shown in Fig.~\ref{Fig-3}. According to Fig.~\ref{Fig-3}, we know the function $G(Q_*)$ fits well with the numerical results.

\begin{widetext}
\begin{center}
\begin{figure}
\includegraphics[width=0.85 \textwidth]{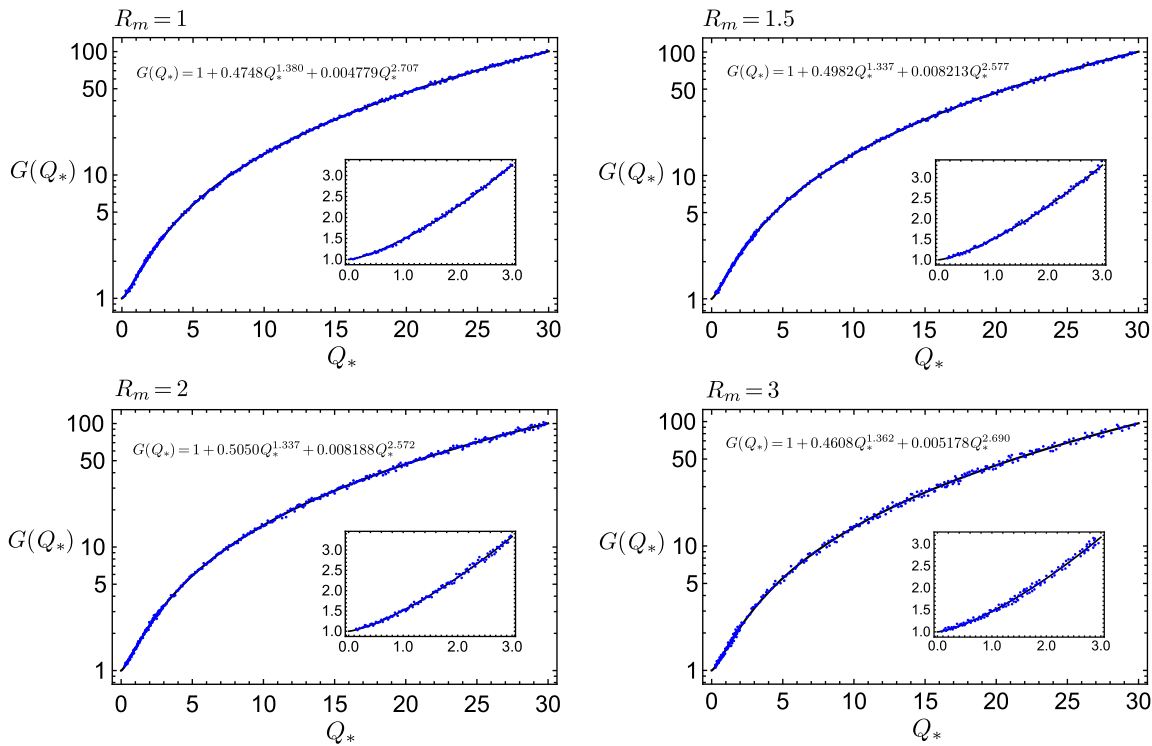}
\caption{The numerical results (blue dots) and analytic fitting function of $G(Q_*)$ (black line) are shown against $Q_*$. In general, the analytic expression $G(Q_*)$ fits well with the numerical results, and the figure shows the growing function $G(Q_*)$ is slightly different for different potentials. At the lower right part of each panel, we zoom in to give more details for small values of $Q_*$.}
\label{Fig-3}
\end{figure}
\end{center}
\end{widetext}

With a given mass ratio $R_{\rm m}$, for every set of initial conditions $(\phi_*, \chi_*)$, we can determine the value of $m_\phi$ and $C_T$ using $N_e=60$ and observational constraints $\mathcal{P}_{\mathcal{R}}=2.2 \times 10^{-9}$ at the end of inflation. And then with the value of $(m_\phi, C_T)$ we can obtain the spectral index and tensor-to-scalar ratio $(n_s, r)$ with Eqs.~\eqref{spectral index} and \eqref{tensor-to-scalar ratio}. Therefore, we have a corresponding $(n_s, r)$ for every set of initial condition $(\phi_*, \chi_*)$. We show the final value of spectral index $n_s$ with initial condition $(\phi_*, \chi_*)$ in Fig.~\ref{Fig-4}. In every panel in Fig.~\ref{Fig-4}, the damping strength $Q_*$ is larger when $(\phi_*, \chi_*)$ are near to original point $(0,0)$. According to the figure we know the strong dissipative effect will render the spectrum blue-tilted, which agree with the conclusion in single-field cases. In all directions, the farthest $(\phi_*, \chi_*)$ from original point represent the case $Q_*=0$ (cold inflation). We also find when $R_m=1$ ($m_\phi=m_\chi$), $n_s$ is the same for all the points $(\phi_*, \chi_*)$ of same distance from original point. However, this property of symmetry no longer exists when $R_m \not =1$. In fact, when $R_m=1$, the background evolution trajectory is a straight line in field space of $(\phi, \chi)$, so there is only one degree of freedom, which is the same as single-field cases. In the following context, we use this case to represent an example of single-field warm inflation for comparison with other cases.

\begin{widetext}
\begin{center}
\begin{figure}
\includegraphics[width=0.85 \textwidth]{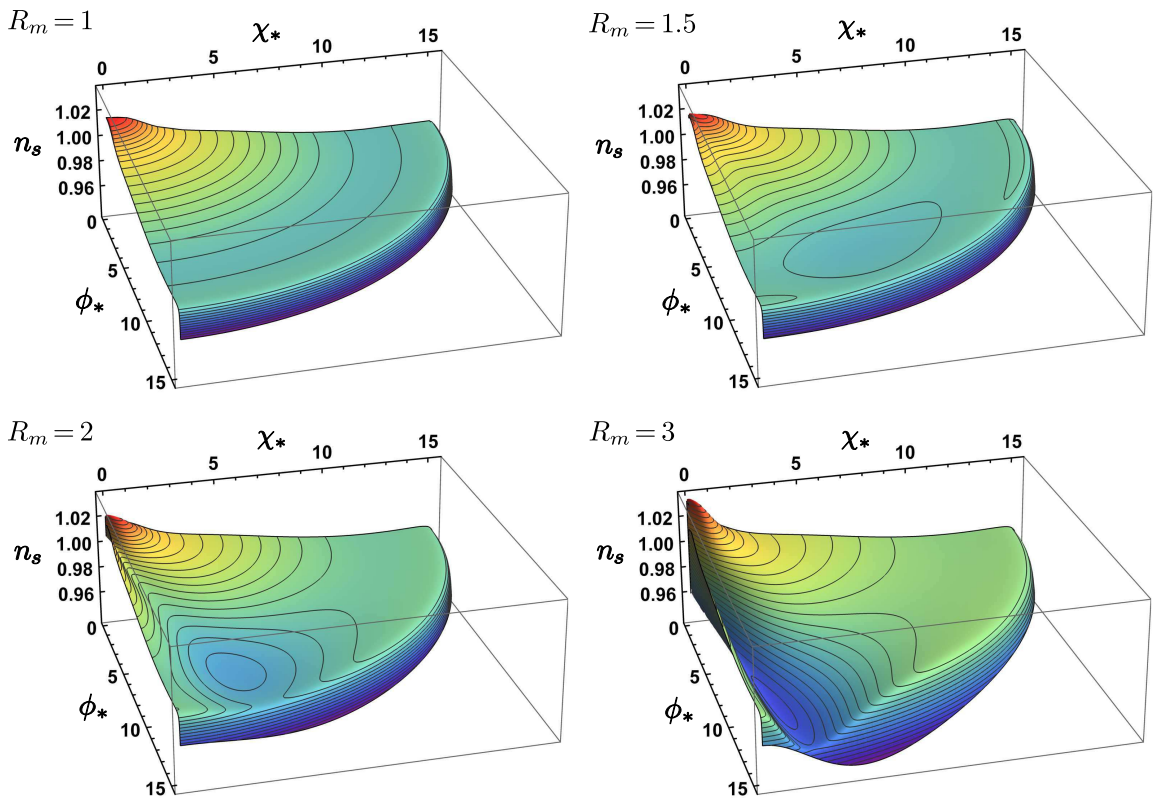}
\caption{The final spectral index $n_s$ are plotted against the initial condition $(\phi_*, \chi_*)$. In each panels, the dissipative ratio $Q_*$ are generally larger when $(\phi_*, \chi_*)$ are closer the the original point $(0,0)$, and the farthest points in each direction denotes $Q_*=0$ (cold two-field inflation). The figure also shows that the dissipative effects will render the power spectrum blue-tilted. Meanwhile, multi-field effects will make $n_s$ take values in a wider range.}
\label{Fig-4}
\end{figure}
\end{center}
\end{widetext}

In order to compare our results with observational data directly, we also show our results in $(n_s, r)$ plane, as illustrated in Fig.~\ref{Fig-5}. For each $R_{\rm m}$, we plot $(n_s, r)$ in the same plot with the allowed contour plots of Planck data ($68\%$ and $95\%$ C.L. results from Planck 2018 TT,TE,EE+lowE+lensing)~\cite{planck 2018}. Note that in the figure we represent $Q_*=0$, $Q_*=1$, and $Q=10$ with thick black lines (the line for $Q_*=0$ is at the top of the blue-shaded region, which represent the cold inflation cases).  As described above, we can treat the $R_{\rm m}=1$ case as a single-field warm inflation example. Comparing $R_{\rm m}=1$ with other values of $R_{\rm m}$, we can conclude that the multi-field effects will make $n_s$ distribute in a wide range for every value of $r$, rather than a point. Besides, the dissipative effects have an impact both on the spectral index $n_s$ and tensor-to-scalar ratio $r$, while the multi-effects have little influence on $r$. According to Fig.~\ref{Fig-5} we know in single-field cases, strong version of warm inflation is disfavored by observation for the dissipation will render the spectrum blue-tilted.  However, warm inflation can happen when $Q_*>1$ in multi-field cases. For all cases we studied In Fig.~\ref{Fig-5}, $n_s$ shows oscillatory features when $Q_*$ changes, and this interesting feature also happen in Ref.~\cite{constraining warm inflation}

We find $Q_*<0.0018$, $r>0.0090$ for $R_{\rm m}=1$, which means only weak versions of warm inflation is allowed by observation, and this is consistent with the previous studies on single-field warm inflation \cite{warm inflation dissipative}. When $R_{\rm m}=1.5$, the dissipative strength $Q_*$ can take values $Q_*<0.0024$, and $r>0.0077$. In case of $R_{\rm m}=2$, $Q_*$ lies in a slight wider range $Q_*<0.0044$, and $r$ takes values $r>0.0057$. For $R_{\rm m}=3$, $Q_*$ can be as large as $100$, which means warm inflation is no longer restricted to weak regimes, and this is very different from single-field cases. In this case, $r$ has a lower bound about $10^{-15}$.

\begin{widetext}
\begin{center}
\begin{figure}
\includegraphics[width=0.85 \textwidth]{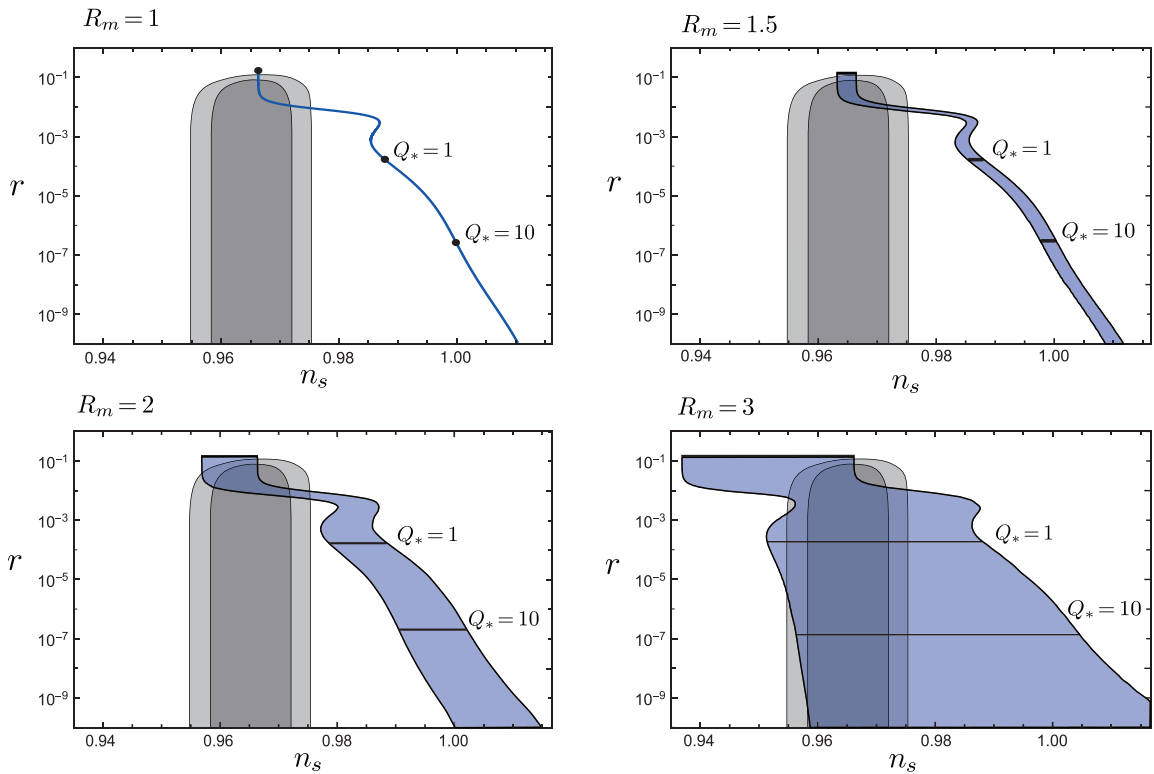}
\caption{Observational predictions (blue regions) of two-field warm inflation models with differential mass ratio $R_{\rm m}$. The gray contours correspond to the $68 \%$ and $95 \%$ C.L. results from Planck 2018 TT,TE,EE+lowE+lensing data. Note that when $R_{\rm m}=1$, the results are same to the single-field case. The black lines at the top of blue regions in each panel denote $Q_*=0$ (cold two-field inflation). The lower right panel shows that strong dissipative warm inflation is observationally favored due to the multi-field effects in warm inflation. }
\label{Fig-5}
\end{figure}
\end{center}
\end{widetext}

Another potential discriminator between different inflationary models is the non-Gaussianity produced during inflation. The nonlinear parameter $f_{\text{NL}}$ is given by,

\begin{equation}
f_{\text{NL}} = f_{\text{NL}} ^{(3)} + f_{\text{NL}} ^{(4)},
\end{equation}
where $f_{\text{NL}} ^{(3)}$ is a slow-roll suppressed term coming from the intrinsic non-Gaussianity in $\delta \phi$ and $\delta \chi$. It has been shown that $f_{\text{NL}} ^{(3)}$ is much less than unity~\cite{non-Gaussianity of}, which is too small to be observed by the current CMB experiment. Therefore, we will concentrate on the second term $f_{\text{NL}} ^{(4)}$, which can be expressed using $\delta N$ formalism~\cite{the next non-Gaussianity},
\begin{equation}
f_{\text{NL}}^{(4)} = \frac{5}{6} \frac{\sum_{IJ} N_{,IJ} N_{,I} N_{,J}}{\left( \sum_I N_{,I}^2 \right)^2}, \label{fnl}
\end{equation}
where $N_{,I} = \partial N/(\partial \varphi_*^I)$, $N_{,IJ} = \partial^2 N/(\partial \varphi_*^I \partial \varphi_*^J)$, and the index $I, J$ run over all of the fields.

We have performed some numerical calculations of $f_{\text{NL}}^{(4)}$ using Eq.~\eqref{fnl}, and obtained the final value of $f_{\text{NL}}^{(4)}$ at the end of inflation. According to our results, the nonlinear parameter $f_{\text{NL}}^{(4)} = 2.3 \times 10^{-3}$ for strong dissipative warm inflation, and $f_{\text{NL}}^{(4)} = 6.9 \times 10^{-3}$ for weak dissipative warm inflation. Our results are of the same order of magnitude as some previous studies on warm inflation and multi-field inflation~\cite{non-gaussianities in two-field, dynamics and non-Gaussianity, spectral running and, non-Gaussianities in}. In Fig.~\ref{Fig-6}, taking $R_{\text{m}} = 2$, $5$ and $Q_* = 0.1$, $10$ as examples, we give the evolution of $f_{\text{NL}}^{(4)}$ from Hubble exit until the end of inflation. As illustrated in Fig.~\ref{Fig-6}, $f_{\text{NL}}^{(4)}$ grows sharply when the heavy field decay to zero, corresponding to the turn of trajectory in field space, but then decrease. After this moment, there is only one effective field and the nonlinear parameter become slow-roll suppressed, which is the same as single-field inflation. We can also conclude from Fig.~\ref{Fig-6} that multi-field inflation do not necessarily produce large non-Gaussianity, and the mass ratio $R_m$ does not have a significant impact on the final nonlinear parameter $f_{\text{NL}}^{(4)}$.

\begin{widetext}
\begin{center}
\begin{figure}
\includegraphics[width=0.9 \textwidth]{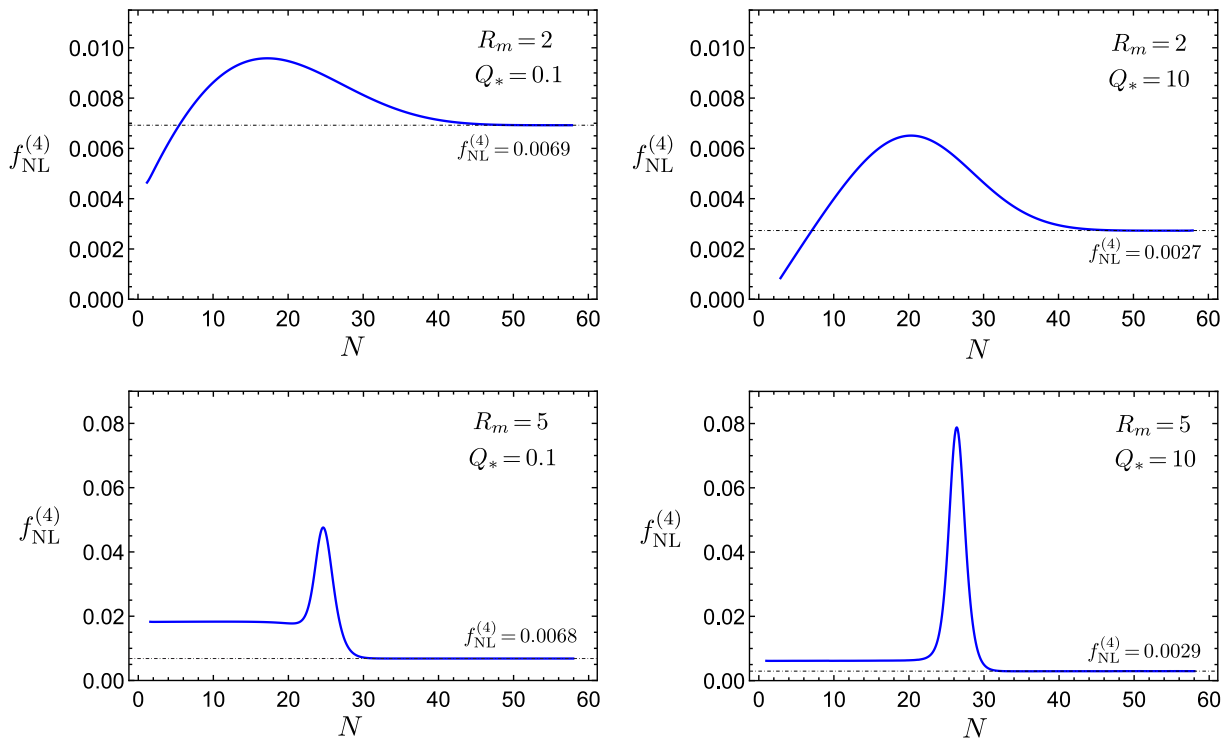}
\caption{The evolution of nonlinear parameter $f_{\text{NL}}^{(4)}$. We take $R_m=2$, $5$ and $Q_* = 0.1$, $10$, and the values of $f_{\text{NL}}^{(4)}$ at the end of inflation are indicated in each panel. In our numerical examples, turns of trajectories in field space all occur between $N=10$ and $N=30$.}
\label{Fig-6}
\end{figure}
\end{center}
\end{widetext}

\section{Conclusions}

In this work, we have tried to explore the main features of perturbations in two-field warm inflation, and then use the observational data to constrain our models. We use the two-field quadratic warm inflation model with a linear-$T$ dissipative coefficient as a representative example, and carry out exhaustive numerical simulations to reveal the main features of multi-field warm inflation. Firstly, we derived the full set of equations describing background dynamics and perturbations. Taking into account of the stochastic noise, these are a set of coupled stochastic differential equations (SDE). Then we apply the formula to an example and get the main features of the evolution of perturbations by solving the SDE numerically. We have shown that the curvature perturbation $\mathcal{R} \approx \mathcal{R}_\sigma \approx \mathcal{R}_r$ in super-horizon scales and the isocurvature will decay to zero before the end of inflation. In the following calculation, instead of integrating the SDE to the end of inflation directly, we take a less compute-intensive method to get the power spectrum at the end of inflation. We use a analytic formula Eq.~\eqref{analytic expression} to describe $\mathcal{P}_{\mathcal{R}}$ at horizon crossing, in which the growing function $G(Q_*)$ is obtained by fully numerical method. After horizon crossing, $\delta N$-formalism is used to get the final power spectrum $\mathcal{P}_{\mathcal{R}}$ at the end of inflation. With the observational constraints $\mathcal{P}_{\mathcal{R}}=2.2 \times 10^{-9}$ and $N_e=60$, we obtain a set of $(n_s, r)$ for every initial condition $(\phi_*, \chi_*)$, as illustrated in Fig.~\ref{Fig-4}. At last, we show our results in $(n_s, r)$ plane in Fig.\ref{Fig-5}, and compare the observational predictions of our models with the latest Planck data.

According to Fig.~\ref{Fig-5}, we know that in single-field inflation, warm inflation only occurs in weak dissipative regime, because the strong dissipation will render the power spectrum blue-tilted, which is not compatible with observation. However, the multi-field effects will cause the spectral index $n_s$ take values in a wide range for every $r$, rather than just a point. In this condition, observations are less effective in constraining $n_s$, and when the multi-effects is large enough strong versions of warm inflation can become favored by observation.

The thick black line in each panels of Fig.~\ref{Fig-5} represents $Q_*=0$ (cold inflation), from which we know the multi-effects will not change the tensor-to-scalar ratio $r$ very much. Therefore, the inflationary models ruled out by observation for predicting to large $r$ may not be rescued in the multi-field case.

The existence of isocurvature perturbations will lead to the time evolution of the curvature perturbation $\mathcal{R}$, and this evolution can happen even in post-inflationary era. Therefore, multi-field inflation is generally not predictive, unless an adiabatic limit is reached before the end of inflation. Fortunately, in all the cases we studied, the isocurvature mode of perturbations decay to zero before the end of inflation, as illustrated in Fig.~\ref{Fig-2}. However, Fig.~\ref{Fig-5} shows even in this condition, two-field warm inflation models are not well constrained by the observational results of $(n_s, r)$, especially when the mass ratio $R_{\rm m}$ is large. In this case, other observational predictions such as the running of spectral index can be used to constrain inflationary models, and we will leave this to future study. On the other hand, if two mass scales are widely separated, inflation tends to be dominated by only the light field. Under this circumstance, we will go back to single-field warm inflation.

Compared with single-field cases, slow-roll parameters may become relatively large in multi-field warm inflation. We have to go beyond the slow-roll approximations, or some important features will be lost. Fig.~\ref{Fig-2} shows that the slow-roll parameters $\epsilon$ and $\eta$ have local extremes when the heavy field $\chi$ decay to zero, and this effect becomes more obvious when the mass ratio $R_{\rm m}$ is large. If these extremes occur at horizon crossing by coincidence, the slow-roll corrections will make the analytic expression Eq.~\eqref{analytic expression} less accurate in describing the power spectrum at horizon crossing. As demonstrated by Fig.~\ref{Fig-3}, larger $R_{\rm m}$ will lead to higher residuals when we use the grow function $G(Q_*)$ to fit the data obtained by numerical simulation. A possible way to deal with this problem is that we can integrate the stochastic perturbation equations Eqs.~\eqref{inflaton perturbation equation}, \eqref{radiation perturbation equation} and \eqref{isocurvature perturbation equation} directly to the end of inflation without introducing the analytic formula Eq.~\eqref{analytic expression} at horizon crossing, which is a more compute-extensive method. From this work, we can conclude that even the simplest two-field warm inflation models are much more complicated than single-field cases, and this topic needs further investigations.

Note that in two-field quadratic warm inflation, when $m_\phi=m_\chi$ the background trajectory in field space is a straight line regardless of the initial conditions. Therefore, the isocurvature mode of perturbation have no influence on curvature perturbation according to Eqs.~\eqref{inflaton perturbation equation}, \eqref{radiation perturbation equation} and \eqref{isocurvature perturbation equation}, so the perturbation features is the same to the single-field case, as illustrated in the top left panel in Fig.~\ref{Fig-5}. However, this property no longer exists in other two-field warm inflation models. For example, in two-field quartic inflation with a potential $V=\frac{1}{4} \lambda_\phi \phi^4+\frac{1}{4} \lambda_\chi \chi^4$, the $\lambda_\phi = \lambda_\chi$ case is not equivalent to single-field inflation (we describe more details in Appendix B). Besides, the dissipative coefficient $\gamma$ can be different for $\phi$ and $\chi$ in general. In this work we use a common $\gamma$ just for simplicity, and this topic needs further study.

\section{Acknowledgements}
This work was supported by the National Natural Science Foundation of China (Grants No. 11575270, No. 11175019, No. 11235003 and No.11605100).

\appendix
\section{Perturbation Equations in Multi-field Warm Inflation}
In this section, we will present some perturbation equations in multi-field warm inflation. For a multi-component system consisting of two scalar field $\varphi_I$ ($I=1,2$) and a radiation fluid, the Einstein equation of the system is given by \cite{perturbation spectra in},

\begin{equation}
G_{\mu \nu} = T_{\mu \nu}^{(\varphi)}+T_{\mu \nu}^{( r )},
\end{equation}
where
\begin{equation}
T_{\mu \nu}^{(\varphi)} \! = \! \Sigma_I \partial_\mu \varphi_I \partial_\nu \varphi_I - g_{\mu \nu} \! \left( \frac {1}{2} \Sigma_I\partial^\lambda \varphi_I \partial_\lambda \varphi_I +V(\varphi) \right),
\end{equation}
\begin{equation}
T_{\mu \nu}^{(r)} \! = \! (\rho_r + p_r) u_\mu u_\nu + p_r g_{\mu \nu},
\end{equation}
where $G_{\mu \nu}$ is the Einstein tensor, and $T_{\mu \nu}^{(\varphi)}$, $T_{\mu \nu}^{(r)}$ are the energy-momentum tensor of scalar fields and radiation fluid. For the radiation fluid, $p_r = \frac{1}{3}
\rho_r$, $\delta p_r = \frac{1}{3} \delta \rho_r$, where $p_r$, $\rho_r$ are the pressure and energy density of the radiation, and $\delta p_r$, $\delta \rho_r$ are their perturbations respectively.

 For simplicity, we will work in spatially-flat gauge ($E=\psi=0$). According to the perturbation equations of Einstein equation, we can represent metric perturbation $A$, $B$ in terms of perturbations of scalar fields and radiation \cite{perturbations in cosmologies},
\begin{equation}
A = \frac{- \delta q_r + \Sigma_I \dot{\varphi_I} \delta \varphi_I}{2 H}, \label{A}
\end{equation}
\begin{multline}
B = - \frac{a}{4 k^2 H^2} (24 H^2 \delta q_r \\
+ (4 \rho_r + \Sigma_I \dot{\varphi_I}^2) (- \delta q_r + \Sigma_I \dot{\varphi_I} \delta \varphi_I) \\
+ 2 H (3 \dot{\delta q_r} + 4 \gamma \Sigma_I \dot{\varphi_I} \delta \varphi_I
+ \Sigma_I \ddot{\varphi_I} \delta \varphi_I - \Sigma_I \dot{\varphi_I}
\dot{\delta \varphi_I})), \label{B}
\end{multline}
where $\delta q_r = a (p_r+ \rho_r) (B + \delta u)$ is the momentum density perturbation of radiation. The variation of scalar field's equation of motion is given by \cite{isocurvature perturbations and},
\begin{multline}
\ddot{\delta \varphi_I} + (3 H + \gamma) \dot{\delta \varphi_I} +
\frac{k^2}{a^2} \delta \varphi_I + \Sigma_J V_{\varphi_I \varphi_J} \delta
\varphi_J + \dot{\varphi_I} \delta \gamma \\
= - (2 V_{\varphi_I} + \gamma \dot{\varphi_I}) A + \dot{\varphi_I} \left( \dot{A} - \frac{k^2}{a}  B \right). \label{phi_I}
\end{multline}

After some adaption, the perturbation of energy and momentum conservation equations of the radiation leads to \cite{two-field warm inflation, xpand: an algorithm},
\begin{multline}
\ddot{\delta q_r} + 7 H \dot{\delta q_r} + 3 \left( 7 H^2 + \dot{H} +
\frac{1}{3} \frac{k^2}{a^2} \right) \delta q_r + \\
\Sigma_I \left( \frac{1}{3} \dot{\varphi_I}^2 \delta \gamma \!+\! (4 H \gamma \!+\! \dot{\gamma}) \dot{\varphi_I}
\delta \varphi_I \!+\! \gamma \ddot{\varphi_I} \delta \varphi_I \!+\! \frac{5}{3}
\gamma \dot{\varphi_I} \dot{\delta \varphi_I} \right) \\
 = \frac{1}{3}
\frac{k^2}{a} \rho_r B - \frac{16}{3} H \rho_r A + \frac{1}{3} \gamma \Sigma_I
\dot{\varphi_I}^2 A - \frac{4}{3} \rho_r \dot{A}. \label{q_r}
\end{multline}

Substituting Eqs.~\eqref{background transform} and \eqref{perturbation transform} into Eqs.~\eqref{phi_I} and \eqref{q_r}, we can get,
\begin{multline}
\ddot{\delta \sigma} + (3 H + \gamma) \dot{\delta \sigma} + \left(
\frac{k^2}{a^2} + V_{\sigma \sigma} - \dot{\theta}^2 \right) \delta \sigma - 2 \dot{\theta} \dot{\delta s} \\
 + 2 \left( \frac{2 \dot{\theta}
V_{\sigma}}{\dot{\sigma}} - \ddot{\theta} \right) \delta s + \dot{\sigma}
\delta \gamma \\
 = - \frac{k^2}{a} \dot{\sigma} B - (\dot{\sigma} \gamma + 2
V_{\sigma}) A + \dot{\sigma} \dot{A}, \label{sigma}
\end{multline}
\begin{multline}
\ddot{\delta s} + (3 H + \gamma) \dot{\delta s} + \left( \frac{k^2}{a^2} +
V_{ss} - \dot{\theta}^2 \right) \delta s \\
+2 \dot{\theta} \dot{\delta \sigma}
- \frac{2 \dot{\theta} \ddot{\sigma}}{\dot{\sigma}} \delta \sigma = 2
\dot{\theta} \dot{\sigma} A, \label{s}
\end{multline}
\begin{multline}
\ddot{\delta q_r} + 7 H \dot{\delta q_r} + 3 \left( 7 H^2 + \dot{H} +
\frac{1}{3} \frac{k^2}{a^2} \right) \delta q_r + \\
\frac{5}{3} \gamma \dot{\sigma} \dot{\delta \sigma}+ (\gamma \ddot{\sigma}+\dot{\gamma} \dot{\sigma}+4H \gamma \dot{\sigma}) \delta \sigma +\frac{1}{3} \dot{\sigma}^2 \delta \gamma - \frac{2}{3} \gamma \dot{\theta} \dot{\sigma} \delta s  \\
 = \frac{1}{3}
\frac{k^2}{a} \rho_r B - \frac{16}{3} H \rho_r A + \frac{1}{3} \gamma {\dot{\sigma}}^2 A - \frac{4}{3} \rho_r \dot{A}. \label{q_rr}
\end{multline}
where
\begin{eqnarray}
V_{\sigma \sigma} &=& \cos^2 \theta V_{\phi \phi} + \sin 2 \theta V_{\phi \chi} + \sin^2 \theta V_{\chi \chi}, \\
V_{s s} &=& \sin^2 \theta V_{\phi \phi} - \sin 2 \theta V_{\phi \chi} + \cos^2 \theta V_{\chi \chi}.
\end{eqnarray}

Substituting Eqs.~\eqref{A}, \eqref{B} into Eqs.~\eqref{sigma}, \eqref{s}, \eqref{q_rr}, and express the equations in terms of $\mathcal{R}_\sigma$, $\mathcal{R}_r$ using $\mathcal{R}_\sigma=H \delta \sigma/\dot{\sigma}$, $\mathcal{R}_r=-H \delta q_r /(p_r+\rho_r)$, we get

\begin{widetext}
\begin{multline}
\ddot{\mathcal{R}_\sigma} + \left( 3H+\gamma+ \frac{4 \rho_r}{3H}+\frac{{\dot{\sigma}}^2}{H}+\frac{2 \ddot{\sigma}}{\dot{\sigma}}  \right) \dot{\mathcal{R}_\sigma} + \left( \frac{k^2}{a^2} + H \gamma - {\dot{\theta}}^2 + \frac{4 \rho_r}{3}+\frac{2 \gamma \rho_r}{3H}+\frac{8 {\rho_r}^2}{9H^2}- \frac{11 \gamma {\dot{\sigma}}^2}{6H} -\frac{\gamma^2 {\dot{\sigma}}^2}{\rho_r}+\frac{4 \rho_r \ddot{\sigma}}{3H \dot{\sigma}} \right) \mathcal{R}_\sigma \\=- \left(\gamma+ \frac{4 \rho_r}{3H} \right) \dot{\mathcal{R}_r} + \left( H \gamma + \frac{4 \rho_r}{3}+\frac{2 \gamma \rho_r}{3H}+\frac{8 {\rho_r}^2}{9H^2}- \frac{11 \gamma {\dot{\sigma}}^2}{6H} -\frac{\gamma^2 {\dot{\sigma}}^2}{\rho_r}+\frac{4 \rho_r \ddot{\sigma}}{3H \dot{\sigma}} \right) \mathcal{R}_r \\+ \frac{2H \dot{\theta}}{\dot{\sigma}} \dot{\delta s} + \left( \frac{6H^2 \dot{\theta}}{\dot{\sigma}}+\frac{2H \gamma \dot{\theta}}{\dot{\sigma}} + \frac{2 H \ddot{\theta}}{\dot{\sigma}} + \dot{\theta} \dot{\sigma}+\frac{2H \dot{\theta} \ddot{\sigma}}{\dot{\sigma}^2} \right) \delta s,
 \end{multline}
 \begin{multline}
\ddot{\mathcal{R}_r}+ \left( -H+\frac{4 \rho_r}{3H}+\frac{{\dot{\sigma}}^2}{H}+\frac{7 \gamma {\dot{\sigma}}^2}{4 \rho_r}\right) \dot{\mathcal{R}_r} + \left(\frac{k^2}{3a^2}+\frac{3 \gamma {\dot{\sigma}}^2}{2H}+\frac{9H \gamma {\dot{\sigma}}^2}{4 \rho_r}+\frac{8 \rho_r {\dot{\sigma}}^2}{9H^2}+\frac{{\dot{\sigma}}^4}{2H^2}+\frac{7 \gamma {\dot{\sigma}}^4}{8 H \rho_r}+\frac{\dot{\sigma} \ddot{\sigma}}{H}+\frac{2\gamma \dot{\sigma} \ddot{\sigma}}{\rho_r} \right) \mathcal{R}_r \\
=\left( \frac{{\dot{\sigma}}^2}{3H}+\frac{5 \gamma {\dot{\sigma}}^2}{4 \rho_r} \right) \dot{\mathcal{R}_\sigma} +\left(\frac{3 \gamma {\dot{\sigma}}^2}{2H}+\frac{9H \gamma {\dot{\sigma}}^2}{4 \rho_r}+\frac{8 \rho_r {\dot{\sigma}}^2}{9H^2}+\frac{{\dot{\sigma}}^4}{2H^2}+\frac{7\gamma {\dot{\sigma}}^4}{8H \rho_r}+\frac{\dot{\sigma} \ddot{\sigma}}{H}+\frac{2\gamma \dot{\sigma} \ddot{\sigma}}{\rho_r} \right) \mathcal{R}_\sigma \\+\left(-\frac{1}{3} \dot{\theta} \dot{\sigma}+\frac{H \gamma \dot{\theta} \dot{\sigma}}{2\rho_r} \right) \delta s,
\end{multline}
\begin{equation}
\ddot{\delta s}+ \left( 3H+\gamma \right) \dot{\delta s}+ \left(\frac{k^2}{a^2}-\dot{\theta}^2+V_{ss} \right) \delta s = -\frac{2 \dot{\theta} \dot{\sigma}}{H} \dot{\mathcal{R}_\sigma}-\frac{4 \dot{\theta} \dot{\sigma} \rho_r}{3H^2} \mathcal{R}_\sigma+\frac{4 \dot{\theta} \dot{\sigma} \rho_r}{3H^2} \mathcal{R}_r.
\end{equation}
\end{widetext}

Note that in case of $\gamma=C_T T$, we have $\delta \gamma/\gamma = \delta T/T=\delta \rho_r/(4\rho_r)$, and $\delta \gamma$ has been eliminated from perturbation equations. Taking into account of the stochastic noise $\xi$ and change the time variable from cosmic time $t$ to e-foldings $N$, we can obtain the perturbation equations in Section II.

\section{Two-field quadratic inflation with equal masses}

According to background equations Eqs.~\eqref{be1} and \eqref{be2}, in case of $m_\phi = m_\chi = m$, for two-field quadratic inflation we have,
\begin{gather}
\phi ''+(3+3Q-\epsilon) \phi ' + m^2 \phi/H^2 = 0, \\
\chi '' + (3+3Q - \epsilon) \chi ' + m^2 \chi/H^2=0,
\end{gather}

After some adaption, the above equations can be put in the form,
\begin{equation}
(\phi' \chi - \chi' \phi)' + (3+3Q-\epsilon) (\phi' \chi - \chi' \phi) = 0. \label{b3}
\end{equation}
The solution of the above equation is given by,
\begin{equation}
(\phi' \chi - \chi' \phi) = C e^{-\int_0^N (3+3Q-\epsilon) dN},
\end{equation}
where $C$ is an integration constant. During inflation, $Q>0$ and $\epsilon<1$, therefore,
\begin{equation}
\left|\phi' \chi - \chi' \phi \right|<\left|C\right| e^{-\int_0^N 2\, dN} = \left|C\right| e^{-2N}. \label{eq0}
\end{equation}

From Eq.~\eqref{eq0}, we know $\phi ' \chi - \chi ' \phi$ will tend to zero rapidly. When $(\phi/ \chi)' = (\phi' \chi-\chi' \phi)/\chi^2 \sim 0$, the trajectory in field space will become a straight line, which is the same as the single-field case.

However, for two-field quartic inflation $V=\frac{1}{4} \lambda_\phi \phi^4+\frac{1}{4} \lambda_\chi \chi^4$, the effective mass $m_\phi = V_{\phi \phi}$ and $m_\chi = V_{\chi \chi}$ are not constants during inflation, and in general Eq.~\eqref{b3} is not valid even for $\lambda_\phi = \lambda_\chi$. In this case, we cannot reach the above conclusion, and the background dynamics of two-field quartic inflation display more complex behavior.

\end{document}